\documentclass[12pt]{article}
\usepackage{latexsym}
\usepackage{amsmath}
\usepackage{amsfonts}
\usepackage{amssymb}
\usepackage{graphicx}
\usepackage{color}
\begin{document}
\input epsf
\def\be{\begin{equation}}
\def\bea{\begin{eqnarray}}
\def\ee{\end{equation}}
\def\eea{\end{eqnarray}}
\def\d{\partial}
\definecolor{red}{rgb}{1,0,0}
\long\def\symbolfootnote[#1]#2{\begingroup%
\def\thefootnote{\fnsymbol{footnote}}\footnote[#1]{#2}\endgroup}
\renewcommand{\a}{\left( 1- \frac{2M}{r} \right)}
\newcommand{\dm}{\begin{displaymath}}
\newcommand{\edm}{\end{displaymath}}
\newcommand{\com}[2]{\ensuremath{\left[ #1,#2\right]}}
\newcommand{\la}{\lambda}
\newcommand{\eps}{\ensuremath{\epsilon}}
\newcommand{\half}{\frac{1}{2}}
\newcommand{\field}[1]{\ensuremath{\mathbb{#1}}}
\renewcommand{\l}{\ell}
\newcommand{\bl}{\left(\l\,\right)}
\newcommand{\normljk}{\langle\l,j,k|\l,j,k\rangle}
\newcommand{\N}{\mathcal{N}}
\renewcommand{\b}[1]{\mathbf{#1}}
\renewcommand{\v}{\xi}
\newcommand{\tr}{\tilde{r}}
\newcommand{\ttheta}{\tilde{\theta}}
\newcommand{\tgamma}{\tilde{\gamma}}
\newcommand{\bg}{\bar{g}}
\renewcommand{\implies}{\Rightarrow}
\newcommand{\z}{\ensuremath{\ell_{0}}}
\newcommand{\temp}{\ensuremath{\sqrt{\frac{2\z+1}{\z}}}}
\newcommand{\twomatrix}[4]{\ensuremath{\left(\begin{array}{cc} #1 & #2
\\ #3 & #4 \end{array}\right) }}
\newcommand{\columnvec}[2]{\ensuremath{\left(\begin{array}{c} #1 \\ #2
\end{array}\right) }}
\newcommand{\e}{\mbox{\textbf{e}}}
\newcommand{\gm}{\Gamma}
\newcommand{\bt}{\bar{t}}
\newcommand{\bphi}{\bar{\phi}}
\newcommand{\m}{\ensuremath{\mathbf{m}}}
\newcommand{\n}{\ensuremath{\mathbf{n}}}
\renewcommand{\theequation}{\arabic{section}.\arabic{equation}}
\newcommand{\newsection}[1]{\section{#1} \setcounter{equation}{0}}
\newcommand{\p}{p}
\newcommand{\tmu}{\tilde{\mu}}
\newcommand{\slthree}{\mathrm{SL}(3,\mathbb{R})}

\vspace{20mm}
\begin{center} {\LARGE Stationary axisymmetric solutions of five dimensional gravity }
\\
\vspace{20mm} {\bf Stefano Giusto and Ashish Saxena}\\ 
\vspace{2mm}
\symbolfootnote[0]{ {\tt giusto@physics.utoronto.ca, ashish@physics.utoronto.ca}} 
Department of Physics,\\ University of Toronto,\\ 
Toronto, Ontario, Canada M5S 1A7.\\
\vspace{4mm}
\end{center}
\vspace{10mm}
\begin{abstract}
We consider stationary axisymmetric solutions of general relativity that asymptote to five dimensional Minkowski space. It is known that this system has a hidden SL(3,R) symmetry. We identify an SO(2,1)  subgroup of this symmetry group that preserves the asymptotic boundary conditions. We show that the action of this subgroup on a static solution generates a one-parameter family of stationary solutions carrying angular momentum.  We conjecture that by repeated applications of this procedure one can generate all stationary axisymmetric solutions starting from static ones.
As an example, we derive the Myers-Perry black hole starting from the Schwarzschild solution in five dimensions. 
\end{abstract}

\thispagestyle{empty}
\newpage
\setcounter{page}{1}

\newsection{Introduction}\setcounter{equation}{0}
Einstein equations on spaces possessing Killing vectors typically have hidden symmetries~\cite{maison, geroch} that
are of both theoretical and technical importance. From the theoretical point of view, a discrete subset
of these symmetries survives as an exact symmetry group of string theory: these are the famous duality groups of string theory~\cite{duality}. On the technical side, these symmetries can be important tools for the solution of Einstein equations, or for the generation of
new solutions from known ones~\cite{solutiongenerating}.  The knowledge of exact solutions~\cite{stephani} of General Relativity has played a crucial role in our understanding of space-time.

In this article we consider Einstein gravity in five dimensions. It has been been known for a long time~\cite{maison} that if one restricts to stationary solutions with at least one spatial Killing vector,  Einstein equations are invariant under an $\mathrm{SL}(3,\mathbb{R})$ symmetry group.  A generic element of this group changes the asymptotic behavior of the geometry in an uncontrolled way and thus cannot be given any physical interpretation. It is thus important to identify the
subgroup of  $\mathrm{SL}(3,\mathbb{R})$ that preserves the asymptotic limit.

For five dimensional space-times
of the Kaluza-Klein type, i.e. asymptotic to $\mathbb{R}^{3,1}\times S^1$, the relevant subgroup is
 $\mathrm{SO}(2,1)$. The physical meaning of the action of this subgroup on a given geometry is clear: one generator of  $\mathrm{SO}(2,1)$ adds KK electric charge along the $S^1$, another adds KK magnetic charge and the last one adds NUT-charge. This $\mathrm{SO}(2,1)$ symmetry has
 been applied as a solution generating technique in \cite{rasheed,larsen}.

In this paper we consider five dimensional asymptotically flat space-times, i.e. geometries asymptotic to
$\mathbb{R}^{4,1}$. We restrict ourselves to stationary and axisymmetric solutions, i.e. solutions
having one time-like and two space-like commuting Killing vectors, with the latter corresponding to rotations in two orthogonal planes. One can then apply the results of \cite{maison}, and conclude that Einstein equations restricted to such solutions have an $\mathrm{SL}(3,\mathbb{R})$ symmetry.\footnote{Because of the presence of two space-like Killing vectors, rather than one, the symmetry group is actually larger than $\mathrm{SL}(3,\mathbb{R})$, and perhaps related to the infinite dimensional Geroch group. Though we do not work out  the full symmetry group in this paper, we make use of the presence of the second Killing vector in the following.} To the best of our knowledge, this symmetry has not been exploited as a systematic solution generating technique in this case. One reason might be that it is not immediately obvious how to identify the subgroup
of  $\mathrm{SL}(3,\mathbb{R})$ which preserves $\mathbb{R}^{4,1}$ asymptotics. The starting observation of this paper is that, if one chooses an appropriate combination of the two space-like Killing vectors, the
appropriate subgroup of $\mathrm{SL}(3,\mathbb{R})$ is again isomorphic to $\mathrm{SO}(2,1)$.
More precisely, if $G$ is the subgroup preserving  the $\mathbb{R}^{3,1}\times S^1$ asymptotic form, then the subgroup preserving five dimensional asymptotic flatness is $D^T G D$, where $D$ is an $\mathrm{SL}(3,\mathbb{R})$ matrix that, essentially, converts a $\mathbb{R}^{4,1}$ asymptotically flat solution into a solution having  $\mathbb{R}^{3,1}\times S^1$ boundary conditions. We thus have an $\mathrm{SO}(2,1)$
group of transformations that acts on five dimensional asymptotically flat solutions. How do these transformations change the physical properties of the geometry?  We will see that one effect of these transformations is to add angular momentum. The $\mathrm{SO}(2,1)$ transformations acting on asymptotically $\mathbb{R}^{4,1}$ solutions provide a 
 generating technique that, starting from a static axisymmetric solution generates a stationary solution carrying angular momentum.\footnote{It should be noted that for a generic vacuum solution there is more than one way to add the same angular momentum. For a heuristic example, one can consider multi black hole solutions. A given total angular momentum can be distributed over the individual horizons in multiple ways. One application of the $\mathrm{SO}(2,1)$ transformation will produce one particular configuration among the set of all possible configurations.} One might also make use of the fact that, for axisymmetric solutions, one has two independent space-like Killing vectors. One can perform an $\mathrm{SO}(2,1)$ transformation with respect to the first Killing vector, followed by another $\mathrm{SO}(2,1)$
transformation with respect to the second Killing vector, and so on. In general one will generate a new solution after each step. As an example, one can start from the five dimensional Schwarschild black hole and generate, in two steps, the Myers-Perry geometry with arbitrary angular momenta.

We conjecture that this solution generating technique could be used to generate the most general five dimensional axisymmetric stationary solution, using static solutions as a starting point.
Static solutions in five dimensions with two axial symmetries are well understood~\cite{emparanreall}.  They are completely specified by two independent harmonic functions on three dimensional flat space. The sources of these harmonic functions, called rods, are one dimensional and lie along an axis of the three dimensional flat space. To construct a static solution corresponding to some distribution of rods, it is enough
to solve Laplace equation in three dimensions with the given sources. The situation for non-static, stationary axisymmetric solutions is more involved. The generalization of the method of~\cite{emparanreall} to this case was found in~\cite{harmark}.  This
generalized method provides a nice classification of stationary solutions: a solution is specified by a configuration of rods and, corresponding to each rod, a (normalized) vector in the three-dimensional vector space spanned by the three Killing vectors (one time-like and two space-like). We will refer to the vector associated to each rod as the ``rod orientation''. Given a particular solution one can infer the rod structure. However, unlike the static case, there is no direct way to reconstruct the full solution from the knowledge of the rod structure alone.   In the general stationary case, Laplace equation appearing in the static case is replaced by a non-linear system of differential equations, whose general solution is not known. It is thus of practical importance to have a technique that generates stationary solutions from static ones. We will show that the $\mathrm{SO}(2,1)$ transformations we have described above act in a natural way on the rod structure of the solution: they do not change the number or the position of the rods, but modify the rod orientations. We thus conjecture that, starting with a static solution specified by some configuration of rods, one can generate by an
appropriate sequence of $\mathrm{SO}(2,1)$ transformations, a stationary solution corresponding to the most general orientation of the rods. The larger the number of rods underlying the starting static solution, the larger is the number of steps needed to generate the most general stationary solution with the same number of rods. If this conjecture is true then one has, at least in principle, a technique to generate the most general stationary axisymmetric solution in five dimensions. A similar result is known to be true for four dimensional axisymmetric stationary solutions as was conjectured by Geroch~\cite{geroch} and proved by Ernst and Hauser in~\cite{ernst}. By using the action of the Geroch group, a technique to add angular momentum to static solutions in four dimensions was found in~\cite{clementold}.

The paper is structured as follows. In section 2 we review the $\mathrm{SL}(3,\mathbb{R})$ symmetry of
five dimensional Einstein equations with (at least)  one space-like and one time-like Killing vector and how an $\mathrm{SO}(2,1)$ subgroup of this symmetry group can be used to generate solutions asymptotic to $\mathbb{R}^{3,1}\times S^1$. We then explain that a conjugate subgroup acts on solutions asymptotic to $\mathbb{R}^{4,1}$ and that this subgroup adds angular momentum to a solution. In section 3 we work out in detail the action of these $\mathrm{SO}(2,1)$ transformations on a general static asymptotically flat axisymmetric solution in five dimensions: we compute the asymptotic form of the transformed geometry and its
conserved charges (mass and angular momenta); we also determine how an $\mathrm{SO}(2,1)$
transformation changes the rod structure of a static solution. In section 4 we use a sequence of
$\mathrm{SO}(2,1)$ transformations to generate the Meyrs-Perry solution with two independent angular momenta starting with the five-dimensional Schwarschild black hole. This provides an example of how a sequence of $\mathrm{SO}(2,1)$ transformations can generate the most general stationary solution,
with some fixed number of rods, starting from a static solution. In section 5 we extrapolate from this example and propose a set of conjectures leading to the assertion that the most general axisymmetric stationary solution in five dimensions may be connected to a static solution by a sequence of $\mathrm{SO}(2,1)$ transformations.

\newsection{The $\mathrm{SL}(3,\mathbb{R})$ action}

We start with a brief review of~\cite{maison} where it was shown that there is a ``hidden'' $\slthree$ symmetry of five dimensional Einstein equations restricted to solutions with two commuting Killing vectors. To make the symmetry manifest, the Einstein equations can be rewritten as an $\slthree$ sigma model coupled to three dimensional gravity. This symmetry is responsible for the enhancement of string U-duality groups when considering any of the superstring theories reduced down to three dimensions. These special properties arise because in three dimensions the gauge fields coming from dimensional reduction can be dualized to scalars and the additional transformations mix these dualized vectors with the other scalars coming from the dimensional reduction. In the following we will restrict the discussion to the case with one time-like and two space-like Killing vectors.
  
\subsection{GR as an $\mathrm{SL}(3,\mathbb{R})$ $\sigma$-model}

Consider a stationary solution of five dimensional Einstein gravity with a space-like Killing vector 
${\partial\over \partial \xi^1}$. The solution can be written in the form
\be
ds^2_5 = \lambda_{ab} (d\xi^a+{\omega^a}_i dx^i)(d\xi^b+{\omega^b}_j dx^j)+{1\over \tau} ds^2_3
\label{dec}
\ee
where $a,b=0,1$ and $\xi^0\equiv t$. $ds^2_3$ is a metric on the 3D space with coordinates $x^i$ ($i=1,2,3$); $\lambda_{ab}$ and $\omega^a_i dx^i$ are functions and 1-forms on this space, and we have defined
\be
\tau=-\mathrm{det}\lambda_{ab}
\ee 
The 1-forms $\omega^a$ can be dualized to scalars, $V_a$, as
\be
d V_a = -\tau \lambda_{ab} *_3 d\omega^b
\label{dual}
\ee
where $*_3$ is performed with the metric $ds^2_3$. As shown in \cite{maison}, the integrability of this equation is guaranteed by the Einstein equations for the metric in Eq. (\ref{dec}). Eq. (\ref{dual}) defines $V_a$ up to arbitrary constants, that can be fixed by imposing some natural boundary conditions at the asymptotic infinity.
The set of scalars $\lambda_{ab}$ and 
$V_a$ can be organized in the following $3\times 3$ symmetric unimodular matrix
\be
\chi=\begin{pmatrix}
\lambda_{ab} -{1\over \tau} V_a V_b & {1\over \tau} V_a\cr {1\over \tau} V_b & -{1\over \tau}
\end{pmatrix}
\ee
In terms of the matrix $\chi$, the equations of motions can be written in the compact form
\bea
&&d *_3(\chi^{-1} d\chi)=0 \label{chieq}\\
&&R^{(3)}_{ij}={1\over 4}\mathrm{Tr} (\chi^{-1}\partial_i \chi\,\chi^{-1}\partial_j \chi)
\label{eom}
\eea
where $R^{(3)}_{ij}$ is the Ricci tensor for the metric $ds^2_3$. 

Eq. (\ref{chieq}) can be interpreted as guaranteeing that the matrix of two forms $*_{3} \chi^{-1}d\chi$ is integrable. We can exploit this by defining the matrix of one forms $\kappa$:
\be
\chi^{-1} d\chi = *_3 d\kappa
\label{kappadef}
\ee
$\kappa$ is defined up to the addition of a matrix of closed 1-forms: this ambiguity can be resolved by imposing 
suitable boundary conditions at asymptotic infinity. We will specify a natural set of boundary conditions on $\kappa$ in the following sections.
It can be shown that some components of the matrix equation (\ref{kappadef}) defining $\kappa$ reduce to the duality equations
(\ref{dual}); in this way one can prove the following useful observation
\be
\omega^0= -{\kappa^0}_2\,,\quad \omega^1 =-{\kappa^1}_2
\label{kappaomega}
\ee

Rewriting the Einstein equations in terms of the matrix $\chi$ has the advantage of making manifest the classical symmetries  
of the system. Indeed, consider the following transformation
\bea
\chi\to \chi'=N\chi N^T\,,\quad ds^2_3\to ds^2_3\quad \mathrm{with}\quad N\in \mathrm{SL}(3,\mathbb{R})
\label{5Dtrans0}
\eea
This transformation preserves the fact that $\chi$ is symmetric and unimodular and leaves the equations of motion (\ref{chieq},\ref{eom})
invariant. Thus, given a five dimensional solution corresponding to the set of data $(\chi,ds^2_3)$, 
the geometry corresponding to $(\chi',ds^2_3)$ is another solution of Einstein equations. To reconstruct the geometry from the data
$(\chi', ds^2_3)$ one has to solve the duality equations (\ref{dual}) to compute the transformed 1-forms $\omega^a$. This problem can be reduced to a purely algebraic one by looking at the matrix $\kappa$. From the definition (\ref{kappadef}) it is clear that under (\ref{5Dtrans0}), $\kappa$ transforms as
\be
\kappa\to \kappa'=(M^T)^{-1}\kappa M^T
\label{krot}
\ee
One can then use the fact that the 1-forms $\omega^a$ sit inside the matrix $\kappa$, as specified in Eq. (\ref{kappaomega}), to extract
from $\kappa'$ the 1-forms $\omega^a$ for the transformed solution.

It is important to note that a general $\slthree$ transformation will not preserve the asymptotic structure of the metric. In any practical application
of the transformations one is usually interested in keeping the asymptotic structure fixed. In the following subsections we examine some boundary conditions of interest and identify the appropriate ``isotropy'' subgroups.

\subsection{Boundary Conditions I: $\mathbb{R}^{3,1} \times S^{1}$}

We first consider solutions which asymptotically approach $\mathbb{R}^{3,1} \times S^{1}$. This is the case of Kaluza-Klein gravity and has been extensively studied in the literature. The asymptotic behaviour of the five dimensional metric is assumed to be described by 
\be
ds^2 = -dt^2 + dr^2 +r^2 (d\theta^2+ \sin^2\theta d\phi^2) + (dx_5)^2
\ee
Here $x_5$ parametrizes an $S^{1}$ of fixed radius. It is straightforward to work out the matrix $\chi$ for this metric by choosing $\xi^{0}=t,\ \xi^{1} = x_5$. The gauge potentials $\omega^{a}$ and their duals $V_a$ are zero. The $\chi$ is given by
\be
\eta_4  \equiv \begin{pmatrix}-1&0&0\cr 0&1&0\cr0&0&-1\end{pmatrix}
\ee 
This indicates that for an asymptotically $\mathbb{R}^{3,1} \times S^{1}$  solution, the asymptotic behaviour of $\chi$ is
\be
 \chi\rightarrow \eta_4 \label{asympchi4}
\ee 
The subgroup of  $\mathrm{SL}(3,\mathbb{R})$ that preserves the boundary condition on $\chi$ is composed of those matrices which satisfy
\be
N\eta_4 N^T =\eta_4
\label{so21}
\ee
i.e. $\mathrm{SO}(2,1)$. These transformations can then be used to transform from one given solution to another~\cite{rasheed}. The only complication with this procedure is that all $\chi$ satisfying Eq. (\ref{asympchi4}) may not be asymptotically Minkowskian. In particular, they could have a NUT charge or a KK-monopole charge. A general SO(2,1) transformation will connect such solutions to the ones which are asymptotically flat in the usual sense. In order to get physically interesting solutions one has to restrict the parameters of the SO(2,1) transformation. For more details on the issue of NUT charge elimination in this case, see~\cite{rasheed}. 

\subsection{Boundary Conditions II: $\mathbb{R}^{4,1}$}
\label{5dflatsec}

The fact that the SO(2,1) transformations described above, can be used to generate solutions of Kaluza-Klein theory is well known. One of the main observations of this paper is that one can extend the formalism to the case of solutions which are asymptotically five dimensional Minkowski space, i.e. $\mathbb{R}^{4,1}$. Furthermore, this generalization when restricted to the case of axisymmetric stationary five dimensional solutions seems to be even more powerful than the known case of Kaluza-Klein gravity. The first step in this direction is to show that there is a particular choice of the space like Killing vector $\xi^{1}$ such that the $\chi$ matrix is asymptotically constant. Though one can choose $\xi^{1}$ to be any space like linear combination of the Killing vectors, a generic choice will lead to $\chi$ being non-constant on the asymptotic $S^3$. The isotropy subgroup of $\slthree$ which will preserve such a function will typically be trivial. As an example consider five dimensional Minkowski space
\be
ds^2 = -dt^2 + dr^2 + r^2 (d\theta^2 + \sin^2\theta d\phi^2 + \cos^2\theta d\psi^2) \label{flat5d}
\ee
and choose $\xi^{0} =t,\ \xi^{1} = \ell \psi$ where $\ell$ is some arbitrary length scale. Then we find
\be
\chi=\left(
\begin{array}{lll}
 -1 & 0 & 0 \\
 0 & \frac{r^2 \cos ^2\theta }{\ell^2} & 0 \\
 0 & 0 & -\frac{\ell^2 \sec ^2\theta }{r^2}
\end{array}
\right)
\ee
One finds that only the identity preserves the above structure. To uncover a richer structure we must choose $\xi^1$ appropriately. It turns out that there are two choices. We can either take $\ell(\psi+\phi)$ or $\ell(\psi -\phi)$. In both cases $\chi$ becomes constant asymptotically. For concreteness let us make the first choice
\be
\xi^1=\ell(\psi+ \phi)
\label{x1def}
\ee
where as before, $\ell$ is an arbitrary parameter of dimension length. We also define $\phi_{-}= \psi- \phi$. Then flat space in Eq. (\ref{flat5d}) can be brought to the form in Eq. (\ref{dec}) with the following result
\bea
&&\lambda_{00}= -1,\ \lambda_{11} = \frac{r^2}{4 \ell^2},\ \lambda_{01}=0,\ \omega^0=0,\ \omega^{1}=\ell \cos 2\theta d\phi_{-}, \ \tau= \frac{r^2}{4\ell^2},\nonumber \\ && \ ds_3^2 = \frac{r^2}{4\ell^2} \left[dr^2 + r^2 d\theta^2 + r^2 \sin^2\theta \cos^2\theta d\phi_{-}^2 \right]
\label{asy}
\eea
Solving Eq. (\ref{dual}) the dual twist potentials can be found. One finds
\be
V_{0}=0,\ V_{1} = \frac{r^2}{4\ell^2}
\ee
The matrix $\chi$ for flat space then becomes
\be
\chi= \begin{pmatrix}-1&0&0\cr0&0&1\cr0&1& -\frac{4\ell^2}{r^2}
\end{pmatrix} 
\ee
In the large $r$ limit $\chi$ becomes
\be
\chi\stackrel{r\rightarrow \infty}{\longrightarrow} \eta_5\equiv \begin{pmatrix}-1&0&0\cr0&0&1\cr0&1&0
\end{pmatrix}
\label{b5}
\ee
Thus an asymptotically $R^{4,1}$ solution will have a $\chi$ matrix approaching $\eta_{5}$ as $r\rightarrow \infty$. The twist potentials $V_a$ are determined only upto additive constants by Eq. (\ref{dual}). By an appropriate choice of these constants the asymptotic limit of the matrix $\chi$ for any asymptotically flat solution can always be brought to the above form.

In the following we will also need to have control over the subleading correction to the matrix $\chi$ at large $r$. 
We will assume the following asymptotic form for $\chi$
\be
\chi = \eta_5\Bigl[1-{\delta\chi\over r^2}+O\Bigl({1\over r^4}\Bigr)\Bigr]
\label{chiasy}
\ee
where $\delta\chi$ is a traceless 3 by 3 {\it constant} matrix. While, in principle, $\delta\chi$ could be a function of $\theta$, 
it's not hard to see that if $\delta\chi$ depended on $\theta$, then the matrix $\kappa$, defined through Eq. (\ref{kappadef}), would have
components that go as $\log r $ at infinity;  this would presumably lead to a non-asymptotically flat geometry. It thus seems that the assumption (\ref{chiasy}) is a reasonable one if we restrict to geometries that are asymptotically $R^{4,1}$, and is moreover satisfied in all the cases we have considered. From the above form of $\chi$, and the fact that the asymptotic limit of $ds^2_3$ is the one given in (\ref{asy}), one finds that the matrix of 1-forms $\kappa$ goes to
\be
\kappa= -{\delta\chi\over 4\ell} \cos2\theta d\phi_-+O\Bigl({1\over r^2}\Bigr)
\label{kappaasy}
\ee
We thus conclude that asymptotic flatness implies the form (\ref{chiasy}) and (\ref{kappaasy}) for $\chi$ and $\kappa$.

We now want to find the group of transformations that preserves asymptotic flatness. This coincides with the isotropy subgroup of $\slthree$ which preserves $\eta_5$ i.e. the subgroup $\{M \in\slthree | M\eta_5 M^{T} =\eta_5\}$. The matrix $\eta_5$ is related to $\eta_4$ by an $\mathrm{SL}(3,\mathbb{R})$ matrix $D$ 
\be
\eta_5 = D^T\eta_4 D\,, \quad D=\begin{pmatrix}1&0&0\cr 0& {1\over \sqrt{2}}& {1\over \sqrt{2}}\cr  0& -{1\over \sqrt{2}}& {1\over \sqrt{2}}
\end{pmatrix}
\ee
Using this fact, we see that the subgroup preserving the $\mathbb{R}^{4,1}$ boundary condition i.e. $\eta_5$ is again isomorphic to $\mathrm{SO}(2,1)$, and consists of matrices of the form
\be
M=D^T N D
\label{5Dso21}
\ee
where $N$ satisfies (\ref{so21}).  We also note that a transformation $M$ preserves the asymptotic limits
(\ref{chiasy}) and (\ref{kappaasy}):
\bea
&&\chi\to \chi'=M\chi M^T = \eta_5\Bigl[1-{\delta\chi'\over r^2}+O\Bigl({1\over r^4}\Bigr)\Bigr]\nonumber\\
&&\kappa\to \kappa' = (M^T)^{-1}\kappa M^T = -{\delta\chi'\over 4\ell} \cos2\theta d\phi_-+O\Bigl({1\over r^2}\Bigr)
\label{chipkappap}
\eea
where $\delta\chi'$ is the constant matrix
\be
\delta\chi'=M\delta\chi M^T
\ee
We will now show that the above form of $\chi'$ and $\kappa'$ are sufficient to guarantee that the $M$-transformed metric is still asymptotically $\mathbb{R}^{4,1}$.  The $M$-transformed metric can be written in the from (\ref{dec}), with $\lambda_{ab}$, $\omega^a$ and $\tau$ replaced by some $\lambda'_{ab}$, $\omega'^a$ and $\tau'$,  that can be derived from $\chi'$ and $\kappa'$.
It easily follows from (\ref{chipkappap}) that, for large $r$,
\be
\lambda'_{00} \approx -1,\ \lambda'_{11} \approx {(\sigma r)^2\over 4 \ell'^2},\ \omega'^1 \approx \ell' \cos 2\theta d\phi_-,\ \tau' =  {(\sigma r)^2\over 4 \ell'^2}
\label{trans1}
\ee
where
\be
\sigma^2 = {(\delta\chi)'_{12}\over 4\ell^2},\ \ell'= {(\delta\chi')_{12}\over 4\ell}
\ee
Note also that the base metric $ds^2_3$ can be written as
\be
ds^2_3 = \frac{(\sigma r)^2}{4\ell'^2} \left[d(\sigma r)^2 + (\sigma r)^2 d\theta^2 + (\sigma r)^2 \sin^2\theta \cos^2\theta d\phi_{-}^2 \right]
\label{trans2}
\ee
By comparing $\lambda_{ab}'$, $\omega'^a$ and $ds^2_3$ given in Eqs. (\ref{trans1}, \ref{trans2}) with the flat space values in Eq. (\ref{asy}) one finds that the following change of 
coordinates is needed to bring the asymptotic limit of the metric to an explicitly flat form:
\be
r'= \sigma r,\ \ell'(\psi'+\phi')=\xi^1,\ (\psi'-\phi')=\phi_-
\ee
We still have to show that the $M$-transformed metric has no terms mixing $t$ with $\xi^1$ and $\phi_-$ at large $r$. 
One finds from (\ref{chipkappap}) that $\lambda'_{01}$, which gives the mixing between $t$ and $\xi^1$, approaches  a constant value
\be
\lambda_{01,\infty}'={(\delta\chi')_{02}\over (\delta\chi')_{12}}
\label{l01as}
\ee
for large $r$. In general this value is non-zero, but can be eliminated by the change of coordinates
\be
t\to t'= t-\lambda_{01,\infty}' \xi^1
\ee
Let us now look at the term mixing  $t'$ with $\phi_-$: it is given by
\be
\omega'^0 + \lambda_{01}' \omega'^1
\label{tphimmix}
\ee
At leading order in $1/ r$ we have, using again (\ref{chipkappap}),
\be
\omega'^0\approx {(\delta\chi')_{02}\over 4\ell} \cos2\theta d\phi_- \,,\quad \omega'^1\approx {(\delta\chi)'_{12}\over 4\ell} \cos2\theta
d\phi_- 
\ee
Substituting these expressions in (\ref{tphimmix}), and using (\ref{l01as}), we see that the mixing between $t'$ and $\phi_-$ 
always vanish at large 
$r$. This concludes the proof  that the $M$-transformed metric asymptotes to $\mathbb{R}^{4,1}$.

To summarize, we have shown that a five dimensional metric which is asymptotically $\mathbb{R}^{4,1}$ has a $\chi$ and $\kappa$ of the form
(\ref{chiasy}) and (\ref{kappaasy}). Conversely, any metric that has a  $\chi$ and $\kappa$ of the form
(\ref{chiasy}) and (\ref{kappaasy}) is asymptotically $\mathbb{R}^{4,1}$. The $\mathrm{SO}(2,1)$ group of transformations generated by 
matrices $M$ of the form (\ref{5Dso21}) preserves the form (\ref{chiasy}) and (\ref{kappaasy}) of $\chi$ and $\kappa$, and thus sends
asymptotically $\mathbb{R}^{4,1}$ geometries into asymptotically $\mathbb{R}^{4,1}$ geometries.

\subsection{Generating stationary solutions with SO(2,1)}
\label{gensec}

The $\mathrm{SO}(2,1)$ group is generated by the three following matrices
\be
N_\alpha =\!\! \begin{pmatrix}\cosh\alpha&\sinh\alpha&0\cr \sinh\alpha& \cosh\alpha& 0\cr 0&0&1\end{pmatrix},\ N_\beta =\!\! \begin{pmatrix}1&0&0\cr 0&\cosh\beta&\sinh\beta\cr0& \sinh\beta& \cosh\beta\end{pmatrix},\ N_\gamma =\!\! \begin{pmatrix}\cos\gamma&0&-\sin\gamma\cr0&1&0\cr \sin\gamma&0& 
\cos\gamma\end{pmatrix}
\label{nabc}
\ee
The group of transformations preserving five dimensional boundary conditions is thus generated by
\be
M_\alpha = D^T N_\alpha D\,,\quad M_\beta = D^T N_\beta D\,,\quad M_\gamma = D^T N_\gamma D 
\label{mabc}
\ee
While we are guaranteed that the geometries generated by the action of these transformations are 
asymptotically flat solutions of the five dimensional Einstein equations, it is not a priori clear what their physical properties are. In particular
one would like to understand if the transformations produce physically new geometries, or geometries that are connected by
diffeomorphisms to the starting solutions. We will show that the action of a two-dimensional subgroup of  $\mathrm{SO}(2,1)$ is
unphysical, leaving a one parameter family of physically relevant transformations.

Consider first the transformation generated by $M_\beta$: It is given explicitly by the diagonal matrix
\be
M_\beta=\begin{pmatrix}1&0&0\cr 0 & e^{-\beta}& 0\cr 0&0&e^{\beta}\end{pmatrix}
\ee
Thus the action of $M_\beta$ on the metric coefficients is the following
\bea
&&\lambda_{00}\to \lambda_{00},\ \lambda_{01}\to e^{-\beta} \lambda_{01},\ \lambda_{11}\to e^{-2\beta} \lambda_{11},\ \tau\to e^{-2\beta}\tau\nonumber\\
&&V_0\to e^{-\beta} V_0,\ V_1\to e^{-2\beta} V_1,\omega^0\to e^{\beta} \omega^0, \ \omega^1\to e^{2\beta} \omega^1
\eea
We see that the action of $M_\beta$ on the metric is equivalent to the action of the diffeomorphism
\be
t\to t, \ \xi^1\to e^{-\beta}\xi^1,\ x^i\to e^\beta x^i
\ee
We conclude that the action of $M_\beta$ is unphysical.

Consider now the following change of coordinates
\be
t\to t + s \xi^1,\ \xi^1\to \xi^1,\ x^i \to x^i
\ee
It effects the metric coefficients as
\bea
&&\lambda_{00}\to \lambda_{00},\ \lambda_{01}\to \lambda_{01}+s \lambda_{00},\ \lambda_{11}\to \lambda_{11}+2 s \lambda_{01}+s^2 \lambda_{00},\ \tau\to \tau\nonumber\\
&&V_0\to V_0 - s,\ V_1\to V_1 + sV_0 -{s^2\over 2},\ \omega^0\to \omega^0 - s \omega^1,\ \omega^1\to \omega^1 
\eea
and thus changes the matrix $\chi$ as
\be
\chi\to S\chi S^T \quad \mathrm{with}\quad S=\begin{pmatrix}1&0&s\cr s & 1& {s^2\over 2}\cr 0&0&1\end{pmatrix}
\ee
$S$ leaves $\eta_5$ invariant, and thus can be written as a combination of $M_\alpha$, $M_\beta$, $M_\gamma$:
\be
S=M_\beta M_\alpha M_\gamma\quad \mathrm{with}\quad \tan\gamma=-{s\over \sqrt{2}},\ \cosh\alpha = {1\over \cos\gamma},\ e^\beta = \cos\gamma
\ee
This shows that the one-dimensional subgroup of $\mathrm{SO}(2,1)$ generated by $S$ also acts trivially on five dimensional solutions.

As announced, we are left with a one-dimensional subgroup of $\mathrm{SO}(2,1)$ that, possibly, generates physically new solutions.
Up to change of coordinates, we can take this subgroup as the one generated by $M_\alpha$. What are the physical properties of the solutions generated by this subgroup? 
In section 3 we will start from an axisymmetric {\em static} solution and work out 
the physical properties of the solutions generated by the action of $M_\alpha$. 
We will show that these solutions are, generically, stationary axisymmetric solutions carrying non-vanishing amounts of angular momenta.
While we have not worked out the action of $\mathrm{SO}(2,1)$ transformations on a general
stationary geometry, we conjecture that a similar result applies also to this more general case: acting with an $\mathrm{SO}(2,1)$ transformation on a stationary axisymmetric geometry generically produces a solutions with different amount of angular momentum.

The transformation (\ref{5Dtrans0}) only uses the fact that $\xi^1=\ell(\psi+\phi)$ is a Killing direction. For axially symmetric solutions, in which
both $\psi$ and $\phi$ are Killing directions, one can apply a similar transformation but with respect to the coordinate ${\tilde\xi}^1={\tilde \ell}(\psi-\phi)$.
More precisely, let $(\chi,ds^2_3)$ be the set of data obtained from the decomposition (\ref{dec}) with $\xi^1 = \ell (\psi+\phi)$. From this set of data we can obtain another set of data $(\tilde\chi, d{\tilde s}^2_3)$ by decomposing the same metric in a form analogous to (\ref{dec}) but with $\xi^1$ replaced by ${\tilde\xi}^1={\tilde \ell}(\psi-\phi)$; this is possible because both $\xi^1$ and $\tilde\xi^1$ are Killing directions. We will denote the operation relating
$(\chi,ds^2_3)$ to $(\tilde\chi,d{\tilde s}^2_3)$ as a ``flip''. One can generate a new solution by acting on the data $(\tilde\chi,d{\tilde s}^2_3)$ with an
 $\mathrm{SL}(3,\mathbb{R})$ transformation of the type (\ref{5Dtrans0}):
 \be
 \tilde\chi\to \tilde\chi'=M \tilde\chi M^T\,,\quad d{\tilde s}^2_3\to d{\tilde s}^2_3
\ee 
where $M$ is again of the form (\ref{5Dso21}). These operations can be iterated:
 \be
(\chi,ds^2_3)\stackrel{M}{\longrightarrow}(\chi',ds^2_3)\stackrel{\mathrm{flip}}{\longrightarrow}(\tilde \chi',(d{\tilde s}_3)^2)\stackrel{M'}{\longrightarrow} 
(\chi'',(d{\tilde s}_3)^2)\stackrel{\mathrm{flip}}{\longrightarrow}\ldots
 \ee 
In general, since a ``flip'' and an $\mathrm{SL}(3,\mathbb{R})$ transformation do not commute, one generates a new metric after every $\mathrm{SL}(3,\mathbb{R})$ transformation. In general, the new metrics differ by the previous ones by the amount of angular momentum they carry, and also by their singularity structure. It is conceivable, though we do not know a general proof of this fact, that by an appropriate sequence of $\mathrm{SL}(3,\mathbb{R})$ transformations one generates stationary solutions that carry any desired amount of angular momentum and are free of singularities. As we will explain in the following, one expects the procedure to stop after a finite number of steps, which depends
 on the structure of the starting solution: after this point, the newly generated solutions are just reparametrizations of the previous ones. 
We will give an explicit example of this method in section \ref{mpsec}, where we show that by a sequence of two  $\mathrm{SL}(3,\mathbb{R})$ transformations one can generate the five dimensional Myers-Perry solution, with arbitrary angular momenta, from the static Schwarzschild five dimensional black hole.

\newsection{Spinning the static generalized Weyl solutions}
\label{staticweylsec}
The most general static solution of five dimensional Einstein gravity with two axial symmetries can be written as~\cite{emparanreall}
\be
ds^2 = - e^{2 U_0} dt^2 + e^{2 U_{1} } d{\phi}^2 + e^{2 U_{2} } d{\psi}^2 + e^{2\nu} (dr^2 + dz^2)
\ee
where $U_I$ ($I=0,1,2$) and $\nu$ are functions of $r$ and $z$ and the variable $r$ is defined in
such a way that
\be
U_0+U_1+U_2=\log r
\ee
The functions $U_I$ satisfy the Laplace equation for the 3D Euclidean metric
\be
ds^2 = dr^2 + r^2 d\gamma^2 + dz^2
\label{fictitious}
\ee
where $\gamma$ is a fictitious coordinate on which nothing is assumed to depend. The sources for 
$U_I$ are localized on the $r=0$ axis and can be described by density functions
$\rho_I(z)$, which for physical solutions are piece-wise constants and take the values $\pm 1,0$. Such
sources are denoted as ``rods'' in the literature. For regular solutions the sources are non-negative, and
satisfy
\be
\sum_{I} \rho_I(z)=1 \quad \forall z
\ee
i.e. the rods are not overlapping and cover the $z$ axis exactly once. The function $\nu$ is determined 
in terms of $U_I$ by the following differential relations
\be
\partial_r \nu = -{1\over 2 r}+{r\over 2} \sum_{I=0}^2[(\partial_r U_I)^2 -(\partial_z U)^2 ]\,,\quad \partial_z \nu = r \sum_{I=0}^2\partial_r U_I \partial_z U_I
\label{nueq}
\ee

Let us perform the change of coordinates
\be
\phi_{\pm} = {\psi} \pm {\phi}, 
\ee
and define
\be
U_{\pm} = \frac{U_1 \pm U_2}{2}
\ee
If we rewrite the metric in the form (\ref{dec}) with respect to the coordinate\footnote{To simplify our equations we take $\ell=1$ in most of this section. The factors of $\ell$ can be restored at the end
by simple dimensional analysis.}
\be
\xi^1= \phi_+
\ee 
we find
\bea
&&\lambda_{00}= - e^{2 U_0} \,,\quad \lambda_{11}=  \frac{1}{2} e^{2U_+} \cosh 2U_{-}\,,\quad \lambda_{01}=0\,,\quad \omega^0=0\,,\quad \omega^1=  -\tanh 2 U_- d\phi_-   \nonumber\\
&&ds^2_3 =  \frac{r^2}{4} d\phi_-^2 + \frac{r}{2} e^{2\nu +U_{0} } \cosh 2 U_- (dr^2 + dz^2)\,,\quad 
\tau=  \frac{r}{2} e^{U_0} \cosh 2 U_-
\label{weyl}
\eea
To compute the matrix $\chi$ we need the potentials $V_a$. From the defining relation (\ref{dual}) and (\ref{weyl}), 
we see that
\be
V_0=0
\ee
and that  $V_1$ satisfies 
\be
d V_1 = r\left( \partial_r U_-  dz - \partial_z U_- dr \right) 
\ee
We note that this implies that
\be
\Bigl(\partial_r^2 + \partial_z^2 -{\partial_r\over r}\Bigr)V_1=0
\label{v1rel}
\ee
Given any function $U$, harmonic with respect to the metric (\ref{fictitious}), let us define its ``dual''
$\tilde U$ as the solution of
\be
d{\tilde U} = r(\partial_r U dz -\partial_z U dr)
\label{ddual}
\ee
Integrability of the equation above is equivalent to the harmonicity of $U$. 
We will show in the next subsection that, in all concrete cases of interest, the explicit form for $\tilde U$ can be easily derived.  In this notation, we thus have
\be
V_1 = {\tilde U}_- + c_1
\label{v1}
\ee
where the constant $c_1$ is fixed by imposing the boundary condition (\ref{b5}).

The matrix $\chi$ can now be computed
\be
\chi=\begin{pmatrix}-e^{2 U_0}&0&0\cr 0 & {e^{-U_0}\over 2}\left(r \cosh 2U_- -  {4 V_1^2\over r \cosh 2 U_-}\right)&
{2e^{-U_0}V_1\over r \cosh 2 U_-}\cr 0& {2e^{-U_0}V_1\over r \cosh 2 U_-}& -{2e^{-U_0}\over r \cosh 2 U_-}
\end{pmatrix}
\ee

We can also derive a general form for the matrix of 1-forms $\kappa$. 
The definition $d\kappa = \chi^{-1}\star_3 d\chi$, implies that the non-vanishing components of $\kappa$
satisfy the following equations
\bea
{d\kappa^0}_{0}&=& -r\left( \partial_r U_0  dz - \partial_z U_0 dr \right)\wedge d\phi_- \\
{d\kappa^1}_{1}&=&d(\omega^1 V_1)- r\left( \partial_r U_+  dz - \partial_z U_+ dr \right)\wedge d\phi_-\\
{d\kappa^1}_{2}&=&-d(\omega^1)\\
{d\kappa^2}_{1}&=&d(\omega^1 V_1^2)+ \left({r\over 2}\partial_r V_1 - V_1\right) dz\wedge d\phi_-- {r\over 2} \partial_z V_1 dr\wedge d\phi_- \\
{d\kappa^2}_{2}&=&-d(\omega^1 V_1)- r\left( \partial_r U_+  dz - \partial_z U_+ dr \right)\wedge d\phi_-
+dz\wedge d\phi_-
\eea
These equations can be solved in terms of the dual functions $\tilde U$, defined in (\ref{ddual}), and 
a new quantity $\tilde V_1$, defined by
\be
d{\tilde V}_1= \left({r\over 2}\partial_r V_1 - V_1\right) dz- {r\over 2} \partial_z V_1 dr 
\label{v1t}
\ee
That the above equation is integrable follows from the relation (\ref{v1rel}) satisfied by $V_1$.
Then the solution for $\kappa$ is
\bea
{\kappa^0}_{0}&=&-{\tilde U}_0 d\phi_-+{c^0}_0 d\phi_-\\
{\kappa^1}_{1}&=&\omega^1 V_1 -{\tilde U}_+ d\phi_-+{c^1}_1 d\phi_-\\
{\kappa^1}_{2}&=&-\omega^1\\
{\kappa^2}_{1}&=& \omega^1 V_1^2 +{\tilde V}_1 d\phi_-+{c^2}_1d\phi_-\\
{\kappa^2}_{2}&=&-\omega^1 V_1 -{\tilde U}_+d\phi_- +z d\phi_-+{c^2}_2 d\phi_-
\label{kappamess}
\eea
and the remaining entries vanish. The constants ${c^i}_j$ will be chosen so that at asymptotic infinity
every element of $\kappa$ is proportional to $\cos2\theta d\phi_-$.

\subsection{Explicit expressions}
\label{explicitsec}
 In the case of five dimensional asymptotically flat solutions, the function $U_2$ must have a semi-infinite rod extending from $-\infty$ to
some finite value of $z$, that we denote by $\p_<$. If we define
\be
\zeta_a = z-a\,,\quad R_a = \sqrt{r^2 + (z-a)^2}
\ee
the potential of this semi-infinire rod is
\be
U_<={1\over 2}\log [R_{\p_<} +\zeta_{\p_<}]
\ee
Similarly, $U_1$ must have a semi-infinite rod extending from $z=\p_>$ ($\p_>>\p_<$) to $+\infty$, whose potential is
\be
U_>={1\over 2}\log [R_{\p_>}-\zeta_{\p_>}]
\ee
In addition to these rods, non-trivial solutions have $N$ finite rods: the $i$-th rod starts at $z=\p_i$ and ends at $z=\p_{i+1}$; its length will be denoted as $L_i=\p_{i+1} -\p_i$;  the potential generated by the 
$i$-th rod is
\be
U_i={1\over 2}\log\Bigl[{R_{\p_i}-\zeta_{\p_i}\over R_{\p_{i+1}}-\zeta_{\p_{i+1}}}\Bigr]={1\over 2}\log\Bigl[{R_{\p_{i+1}}+\zeta_{\p_{i+1}}\over R_{\p_{i}}+\zeta_{\p_{i}}}\Bigr]
\ee
Our conventions are summarized by
\be
\p_1=\p_<\,,\quad \p_i = \sum_{j=1}^{i-1} L_j\,,\quad \p_>=\p_<+ \sum_{j=1}^N L_j
\ee
The finite rods are distributed among the three Killing directions $I=0,1,2$, with no two consecutive rods associated to the same direction. We will denote with the index $i_I$ the set of rods associated to the
direction $I$. Then, 
the harmonic functions for five dimensional asymptotically flat solutions are
\bea
U_0=\sum_{i_0} U_{i_0}\,,\quad 
U_1 = U_>+\sum_{i_1} U_{i_1}\,,\quad U_2 = U_<+\sum_{i_2} U_{i_2}
\eea
By construction, $e^{2U_0+2U_1+2U_2}=r^2$.

Associated to any harmonic function $U$ there is a dual function $\tilde U$, defined in (\ref{ddual}). 
The duals of our ``elementary'' harmonic functions $U_>$, $U_<$ and $U_i$ are (with some convenient 
choice of the arbitrary additive constants)
\be
{\tilde U}_>={R_{\p_>}+z\over 2}\,,\quad 
{\tilde U}_<=-{R_{\p_<}-z\over 2}\,,\quad {\tilde U}_{i}={R_{\p_i}-R_{\p_{i+1}}\over 2}
\ee
Putting the various pieces together, one finds the duals of $U_0$, $U_+$ and $U_-$:
\bea
{\tilde U}_0 &=& \sum_{i_0} {\tilde U}_{i_0}={1\over 2}\sum_{i_0} 
(R_{\p_{i_0}}-R_{\p_{i_0+1}})\\
{\tilde U}_+&=&{1\over 2}[{\tilde U}_{>}+{\tilde U}_{<}+\sum_{i_1} {\tilde U}_{i_1}+\sum_{i_2} {\tilde U}_{i_2}]\\
&=&{1\over 4}[R_{\p_>}-R_{\p_<}+2 z+\sum_{i_1} (R_{\p_{i_1}}-R_{\p_{i_1+1}})+\sum_{i_2} 
(R_{\p_{i_2}}-R_{\p_{i_2+1}})]\\
{\tilde U}_-&=&{1\over 2}[{\tilde U}_{>}-{\tilde U}_{<}+\sum_{i_1} {\tilde U}_{i_1}-\sum_{i_2} {\tilde U}_{i_2}]\\
&=&{1\over 4}[R_{\p_>}+R_{\p_<}+\sum_{\p_{i_1}} (R_{\p_{i_1}}-R_{\p_{i_1+1}})-\sum_{i_2} 
(R_{\p_{i_2}}-R_{\p_{i_2+1}})]
\eea
The above functions satisfy
\be
{\tilde U}_0+2{\tilde U}_+=z
\ee
as a consequence of $U_0+2 U_+ = \log r$.
We will show in the next subsection, where we will work out the asymptotic expansion, that with the above definition of $\tilde U_-$ the appropriate value for the constant $c_1$ in (\ref{v1}) is $c_1=0$, so that the potential $V_1$ is
\be
V_1 = {\tilde U}_- 
\ee
With this $V_1$ the Eq. (\ref{v1t}) for $\tilde V_1$ is solved by
\be
{\tilde V}_1=-{1\over 8}[\zeta_{\p_>}R_{\p_>}+
\zeta_{\p_<} R_{\p_<}+\sum_{i_1} (\zeta_{\p_{i_1}}R_{\p_{i_1}}-\zeta_{\p_{i_1+1}}
R_{\p_{i_1+1}})-\sum_{i_2} (\zeta_{\p_{i_2}}R_{\p_{i_2}}-\zeta_{\p_{i_2+1}}R_{\p_{i_2+1}})]
\ee 
At this point one knows all the terms needed for explicitly computing the matrices $\chi$ and $\kappa$. From the asymptotic expansion worked out in the next section, one finds that the appropriate values for the constants ${c^i}_j$ are
\be
{c^0}_0=0\,,\quad {c^1}_1= -{c^2}_2={1\over 4}[\p_<+\p_>-\sum_{i_1} L_{i_1} + \sum_{i_2} L_{i_2}]\,,\quad {c^2}_1=0
\label{c11}
\ee
\subsection{Asymptotic analysis}
In this subsection we study the asymptotic limit of the exact expressions derived above. We will also apply the transformation $M_\alpha$ to a generic static Weyl solution, and study the asymptotic limit 
of the rotating solution thus generated. We will verify that this solution is asymptotically flat and we will derive a general expression for the mass and angular momentum of this solution. 

To write the metric in an explicitly asymptotically flat form, we introduce the coordinates $\rho$ and 
$\theta$
\be
r={\rho^2\over 2}\sin2\theta\,,\quad z={\rho^2\over 2}\cos2\theta
\ee
For a generic distribution of rods, the harmonic functions $U_I$ have the following asymptotic limit
\bea
e^{2 U_I}&=&f_I\Bigl[1-{2\over \rho^2}\delta_I + {2\over \rho^4}(\delta_I^2 -\epsilon_I\cos2\theta )\\\nonumber
&&
+{4\over \rho^6}\Bigl({\nu_I-\delta_I^3\over 3}+\delta_I \epsilon_I \cos 2\theta - \nu_I \cos^2 2\theta \Bigr)+O\Bigl({1\over\rho^8}\Bigr)\Bigl]
\eea
where
\be
f_0=1\,,\quad f_1 = \rho^2 \sin^2\theta\,,\quad f_2 =\rho^2 \cos^2\theta
\ee
and the parameters $\delta_I$, $\epsilon_I$ and $\nu_I$ are defined in terms of the rod distribution as
follows
\bea
\delta_I &=& \sum_{i_I} [\p_{i_I+1}- \p_{i_I}]-\p_> \delta_{I,1}+\p_<\delta_{I,2}\\
\epsilon_I&=& \sum_{i_I} [\p_{i_I+1}^2- \p_{i_I}^2]-\p^2_> \delta_{I,1}+\p^2_<\delta_{I,2}\\
\nu_I &=&\sum_{i_I} [\p_{i_I+1}^3- \p_{i_I}^3]-\p^3_> \delta_{I,1}+\p^3_<\delta_{I,2}
\eea
Note that these parameters satisfy
\be
\sum_I \delta_I =0\,,\quad \sum_I \epsilon_I =0\,,\quad \sum_I \nu_I=0
\ee
The asymptotic expansions for the dual functions $\tilde U_I$ are
\be
{\tilde U}_I={\tilde f}_I + {\delta_I\over 2}\cos2\theta-{\epsilon_I \over 2\rho^2} \sin^22\theta-{\nu_I\over \rho^4}\cos2\theta \sin^2 2\theta + O\Bigl({1\over \rho^6}\Bigr)
\ee
where
\be
{\tilde f}_0=0\,,\quad {\tilde f}_1 = {\rho^2\over 2}\cos^2\theta\,,\quad {\tilde f}_2 = -{\rho^2\over 2}\sin^2\theta
\ee
and 
\bea
{\tilde V}_1 &=& -{\rho^4\over 16}\cos2\theta  + {\rho^2\over 16}(\delta_2-\delta_1)(1+\cos^2 2\theta) -{\epsilon_2-\epsilon_1\over 16}(2+\sin^2 2\theta)\cos 2\theta\nonumber\\
&& +{\nu_2-\nu_1\over 8 \rho^2}\sin^4 2\theta+ O\Bigl({1\over \rho^4}\Bigr)
\eea

We will also need the large $\rho$ expansion of the function $\nu$: this can be obtained by solving eqs. (\ref{nueq}) perturbatively in $1/\rho$. We find
\be
e^{2\nu} = {1\over \rho^2}\Bigl[1+{\delta_0 + (\delta_1 -\delta_2)\cos2\theta\over \rho^2}+O\Bigl({1\over \rho^4}\Bigr)\Bigr]
\ee

One can choose coordinates in such a way that the static metric satisfies the harmonic gauge. If we write the five dimensional metric as 
$g_{\mu\nu}=\eta_{\mu\nu}+h_{\mu\nu}$, where $\eta_{\mu\nu}$ is the five dimensional Minkowski metric, the harmonic gauge requires that
\be
\partial^\mu \Bigl(h_{\mu\nu}-{1\over 2} \eta_{\mu\nu} \eta^{\sigma\rho} h_{\sigma\rho}\Bigr)=0
\ee
We find that, at the first non-trivial order in the large $\rho$ expansion, this gauge condition is satisfied if we take
\be
\delta_1 = \delta_2
\label{harmonic}
\ee
From the expression of $\delta_I$ given above, one sees that we can satisfy this condition by an appropriate choice of the
$z$ origin, the one for which 
\be
\p_< = -{1\over 2}\sum_{i_0}L_{i_0}-\sum_{i_2}L_{i_2}
\ee
We will assume the condition (\ref{harmonic}) in the following, as this will simplify some of our expressions.

Putting things together, we find that the asymptotic expansions of the non-vanishing components of the $\chi$ and $\kappa$ matrices for the
starting static solution are
\bea
{\chi^0}_0 &=& -1 + 2{\delta_0\over \rho^2}+2 {\epsilon_0 \cos2\theta - \delta_0^2\over \rho^4}\nonumber\\
{\chi^1}_1 &=&  {\epsilon_1-\epsilon_2\over 2 \rho^2}+{3 \delta_0 (\epsilon_1 - \epsilon_2)+4(\nu_1-\nu_2)\cos 2\theta\over 6\rho^4}\nonumber\\
{\chi^1}_2 &=& 1 + {\delta_0\over \rho^2}+{\delta_1^2 -2 (\epsilon_1-\epsilon_2)+ 2 \epsilon_0 \cos2\theta \over 2 \rho^4}\nonumber\\
{\chi^2}_2 &=&-{4\over \rho^2}\Bigl(1+{\delta_0\over \rho^2}\Bigr)
\eea 

\bea
{\kappa^0}_0&=&\Bigl[-{\delta_0\over 2}\cos 2\theta + {\epsilon_0\over 2\rho^2}\sin^2 2\theta\Bigr] d\phi_-\nonumber\\
{\kappa^1}_1&=&{\kappa^2}_2=\Bigl[-{\delta_1\over 2}\cos 2\theta + {\epsilon_1+\epsilon_2\over 4\rho^2}\sin^2 2\theta\Bigr] d\phi_-\nonumber\\
{\kappa^1}_2&=& -\cos2\theta d\phi_-\nonumber\\
{\kappa^2}_1&=&\Bigl[{\epsilon_1 -\epsilon_2\over 8}\cos 2\theta - {\nu_1-\nu_2\over 6\rho^2}\sin^2 2\theta\Bigr] d\phi_-\
\eea
We note that $\chi$ and $\kappa$ have the expected asymptotic expansion: this shows that the additive constants in $V_1$ and $\kappa$ we have chosen in the previous
subsection are the appropriate ones. The large $\rho$ limit of the base metric $ds^2_3$ is
\be
ds^2_3 = {\rho^4\over 16}\sin^2 2\theta d\phi_-^2 + {\rho^2\over 4}\Bigl(1+O\Bigl({1\over \rho^4}\Bigr)\Bigr)(d\rho^2 + \rho^2 d\theta^2)
\ee

We can now apply an $\mathrm{SO}(2,1)$ transformation (\ref{5Dtrans0},\ref{krot}), to generate a rotating asymptotically flat solution. A priori
we can act with any combination of the three matrices $M_\alpha$, $M_\beta$ and $M_\gamma$, but, as we proved in section \ref{gensec}, 
it is sufficient to act with $M_\alpha$: the
metric generated by acting with the most general element of $\mathrm{SO}(2,1)$ is related by change of coordinates to the one generated
with $M_\alpha$.

Consider the metric corresponding to the data
\be
\chi'=M_\alpha \chi M_\alpha^T\,,\quad \kappa'=(M_\alpha^T)^{-1}\kappa M_\alpha^T
\label{5Dtransbis}
\ee
and the same base metric $ds^2_3$. This metric is by construction asymptotically flat but carries some amount of angular momentum, that we would like
to compute.

Let $\lambda'_{ab}$, $\tau'=-\mathrm{det}\lambda'_{ab}$  and $\omega'^a$ denote the transformed values of $\lambda_{ab}$, $\tau$ and $\omega^a$, that can be straightforwardly computed from $\chi'$ and $\kappa'$. The rotating metric is given by
\be
ds'^2_5 = \lambda'_{ab} (d\xi^a+{\omega'^a}_i dx^i)(d\xi^b+{\omega'^b}_j dx^j)+{1\over \tau'} ds^2_3 
\ee 
with $\xi^0=t, \xi^1 = \phi_+$. The $t-\phi_+$ mixing in $ds'^2_5$ goes to a non-vanishing constant value for large $\rho$
\be
\lambda'_{01}\approx \lambda'_{01,\infty}=-{\sinh\alpha\over \sqrt{2}}{8 \cosh^2(\alpha/2)-6\delta_0\cosh\alpha - (\epsilon_1-\epsilon_2)\sinh^2(\alpha/2) \over 
  8 \cosh^4(\alpha/2)-3\delta_0\sinh^2\alpha - (\epsilon_1-\epsilon_2)\sinh^4(\alpha/2)}
\ee 
In order to write the metric in a manifestly flat coordinate system one can perform the following change of coordinates
\be
t\to t'= t - \lambda'_{01,\infty} \xi^1
\ee
By construction, the $t'-\phi_+$ mixing now vanishes at infinity; it can also be checked that there is no
$t'-\phi_-$ mixing at large $\rho$.

Let us now look at the $d\rho^2$ term in $ds'^2_5$:  for large $\rho$, it goes to  $\sigma^2 d\rho^2$, with
\be
\sigma^2 ={8 \cosh^4(\alpha/2)-3\delta_0\sinh^2\alpha - (\epsilon_1-\epsilon_2)\sinh^4(\alpha/2)\over 8}
\ee
which implies that the $\rho$ coordinate has to be rescaled as\footnote{We obviously need $\sigma^2>0$. Since $\delta_0>0$ for physically meaningful solutions, one expects $\sigma$ to become negative for large enough $\alpha$. We should thus restrict the range of $\alpha$ to some finite interval 
$\alpha\in[0,\alpha_c]$, for which $\sigma\ge 0$. The value $\alpha_c$ at which $\sigma$ vanishes corresponds to the point at which the rotating solution becomes extremal.}
\be
\rho\to \rho'=\sigma \rho
\ee
We also determine what identifications should be imposed on the coordinate $\phi_+$, such that the metric is asymptotically flat: at this purpose we should look at the $\phi_+-\phi_-$ mixing (at constant $t'$), for large $\rho$. The relevant part of the metric goes to
\be
ds^2_{\phi_+-\phi_-}\approx \lambda_{++} (d\phi_+^2 + \ell' \cos2\theta d\phi_-)^2
\ee
with $\lambda_{++}$ and $\ell'$ constants. One finds
\be
\ell'={8 \cosh^4(\alpha/2)-3\delta_0\sinh^2\alpha - (\epsilon_1-\epsilon_2)\sinh^4(\alpha/2)\over 8}
\ee
Thus if we insist that $\phi_-$ has periodicity $4\pi$, we should require that $\phi_+$ has periodicity $2\pi \ell'$ (effectively this means that the arbitrary 
scale $\ell$, that we have set to 1 for the starting static solution, has been rescaled to $\ell'$ after the transformation (\ref{5Dtransbis})). We can thus introduce
the angular coordinates
\be
\psi'= {1\over 2}\Bigl({\phi_+\over \ell'}+\phi_-\Bigr)\,,\quad \phi'= {1\over 2}\Bigl({\phi_+\over \ell'}-\phi_-\Bigr)\
\ee 
In terms of the coordinates $t'$, $\rho'$, $\theta$, $\psi'$ and $\phi'$, the asymptotic limit of the metric $ds'^2_5$ explicitly reduces to five dimensional Minkowski space. From the first order deviations from flat space we can compute the mass and angular momenta of the solution:
\bea
M&=&2 \delta_0 \cosh^4(\alpha/2)+\Bigl({3\over 8}\delta_0^2 + {\epsilon_1-\epsilon_2\over 4}\Bigr)\sinh^2\alpha-{\delta_0(\epsilon_1-\epsilon_2)\over 4}\sinh^4(\alpha/2)\\
J_\psi&=&{\sinh\alpha\over \sqrt{2}} \Bigl[\Bigl({9\over 4}\delta_0^2 -(\epsilon_1+ 2 \epsilon_2)\Bigr)\cosh^4(\alpha/2)+{\nu_1-\nu_2\over 12} \sinh^2\alpha\nonumber\\
&&+\Bigl({9\over 32} \delta_0^2 (\epsilon_1-\epsilon_2) -{\epsilon_1 ^2 - 2 \epsilon_2 ^2 + \epsilon_1 \epsilon_2\over 8}-{\delta_0 (\nu_1 -\nu_2)\over 4}\Bigr)\sinh^4(\alpha/2)\Bigr]\\
J_\phi&=&{\sinh\alpha\over \sqrt{2}} \Bigl[\Bigl({9\over 4}\delta_0^2 +(2\epsilon_1+  \epsilon_2)\Bigr)\cosh^4(\alpha/2)-{\nu_1-\nu_2\over 12} \sinh^2\alpha\nonumber\\
&&+\Bigl({9\over 32} \delta_0^2 (\epsilon_1-\epsilon_2) -{-2\epsilon_1 ^2 +  \epsilon_2 ^2 + \epsilon_1 \epsilon_2\over 8}+{\delta_0 (\nu_1 -\nu_2)\over 4}\Bigr)\sinh^4(\alpha/2)\Bigr]
\eea

We note that, even if naively the transformation $M_\alpha$ has added momentum along the ``diagonal'' direction $\phi_+=\psi+\phi$,  the angular momenta along the $\psi$ and $\phi$ rotation planes of the resulting solution are not equal $J_\psi\not = J_\phi$. Only if the starting static configuration is symmetric under the exchange of $\psi$ and $\phi$
(implying that $\epsilon_1=-\epsilon_2$ and $\nu_1 = \nu_2$), the final rotating solution has $J_\psi=J_\phi$. 

\subsection{Rod Structure}
In this subsection, we would like to understand how the transformation (\ref{5Dtrans0}) acts on the rod structure. For this purpose we need to generalize the concept of rod structure from the static case to the more general case of stationary axially symmetric solutions. This has been done in \cite{harmark}, to which we refer for more details and for the proofs of some results used in the following.

Consider a generic five dimensional stationary solution of Einstein gravity, admitting three commuting killing vectors ${\partial\over \partial y^I}$ (in our case $y^I=t,\phi,\psi$). The metric can be written as
\be
ds^2 = G_{IJ} dy^I dy^J + e^{2\nu}(dr^2 + dz^2)
\label{stationary}
\ee
with
\be
\mathrm{det} G = - r^2
\label{detG}
\ee
In view of (\ref{detG}), the kernel of $G$ is non-empty for $r=0$. It is shown in \cite{harmark} that  for $r=0$ and generic values of $z$, $G$
has one null eigenvector. Thus the $z$ axis can be split into (finite or semi-infinite) intervals, the rods, over which the null eigenvector of $G$ is constant; 
at the points of separation between rods the kernel of $G$ is 2-dimensional. 

If we rewrite a general stationary metric (\ref{stationary}) in the form (\ref{dec}), we find that the $G$ matrix, in the basis $y^0=\xi^0=t$, $y^1=\xi^1=\phi_+$, $y^2=\phi_-$, has the form
\be
G=\begin{pmatrix}
\lambda_{ab}  & \lambda_{ac} \omega^c_- \cr \lambda_{bc}\omega^c_- & \lambda_{cd}\omega^c_- \omega^d_-+ {r^2\over 4 \tau}
\end{pmatrix}
\label{gmatrix}
\ee
where the indices $a,b,c,d$ range over $0,1$ and $\omega^a_-$ denotes the $\phi_-$-component of the 1-form $\omega^a$. In deriving the identity above we have used the fact that, as a consequence of (\ref{detG}), the base metric for such a solution always has the form
\be
ds^2_3 = {r^2\over 4} d\phi_-^2 + \tau e^{2\nu} (dr^2 + dz^2)
\ee

We want to study the kernel of $G$ at $r=0$. Consider a 3-dimensional vector $v=\{v^a, v_-\}$; $v$ is a null eigenvector of $G$ ($G v =0$)
if
\be
\lambda_{ab}(v^b + \omega^b_- v_-)=0\,,\quad {r^2\over \tau} v_-=0
\label{nulleigen}
\ee
Depending on the form of $\tau$ as $r\to 0$, one can distinguish three cases.
\begin{itemize}
\item case 1: As $r\to 0$, $\tau\to f(z)$, where $f(z)>0$, in some domain on the $z$ axis. In this case the 2 dimensional matrix $\lambda_{ab}$ is invertible and one needs $v^a= -\omega^a_- v_-$. Choosing $v_-=1$, we see that the kernel of $G$ has dimension 1 and is spanned by the vector
\be
v=\{-\omega^a_-,1\}
\ee 

\item case 2: As  $r\to 0$, $\tau\to r f(z)$, where $f(z)>0$, in some domain on the $z$ axis. In this case $\lambda_{ab}$ has one\footnote{We are assuming that matrix
$\lambda_{ab}$ cannot vanish for any value of $z$. Hence its kernel is at most of dimension 1 for all $z$. This is true for all static solutions, and it seems natural to conjecture
that the same is true for stationary solutions as well, though we do not have a proof of this conjecture.} null eigenvector
$v^a_0$ and the matrix $G$ has a 2-dimensional kernel, spanned by
\be
v=\{-\omega^a_-,1\}\,,\quad {\tilde v}=\{v^a_0,0\}
\ee
As we discussed above, this situation can happen only for isolated points on the $z$ axis, which lie at the separation between two rods.

\item case 3:   $r\to 0$, $\tau\to r^2 f(z)$, where $f(z)>0$, in some domain on the $z$ axis. As in the previous case $\lambda_{ab}$ has a null eigenvector 
$v^a_0$, but now we have to take $v_-=0$, due to the second condition in (\ref{nulleigen}). The kernel of $G$ is generated by the single vector
\be
{\tilde v}=\{v^a_0,0\}
\ee

\end{itemize}
Note that $\tau$ cannot go to zero faster than $r^2$, otherwise the term  ${r^2\over 4 \tau} d\phi_-^2$ in (\ref{weyl}) would diverge and the metric would be singular. 

For the static generalized Weyl solutions described in section (\ref{staticweylsec}), case 1 happens in the interior of a space-like rod, case 3 in the
interior of a time-like rod, and case 2 at the boundary between two rods. 

Consider now acting on a static solution with a transformation (\ref{5Dtrans0}):
this will generate a stationary solution of the form (\ref{stationary}). Let us first note that, because the determinant of the matrix $G$ in Eq. (\ref{gmatrix}) always equals $-r^2$ 
and the base metric $ds^2_3$ does not change under the transformation (\ref{5Dtrans0}), we are guaranteed that the Weyl cooordinates $r$ and $z$
of the stationary solution coincide with the ones of the original static solution. Moreover, we will show in the following that  transformations
(\ref{5Dtrans0}) preserve the form of $\tau$. These two facts guarantee that transformations (\ref{5Dtrans0}) do not alter the number nor the position of the rods 
of the solution; the sole effect of (\ref{5Dtrans0}) on the rod structure is to rotate the null eigenvectors of $G$ corresponding to the various rods. In the following
we will start from a static solution and examine the three cases separately.

\begin{itemize}

\item case 1. Consider the limit $r\to 0$ and $z\in (\p_{i_1},\p_{i_1+1})$ or  $z\in (\p_{i_2},\p_{i_2+1})$, i.e. $z$ lies inside a space-like rod. From the explicit expressions given in section \ref{explicitsec}, one sees that $e^{2 U_1}$ vanishes like $r^2$ and $e^{2 U_2}$ stays finite if $z\in (\p_{i_1},\p_{i_1+1})$, or
viceversa if  $z\in (\p_{i_2},\p_{i_2+1})$. In both cases $\lambda_{11}=(e^{2 U_1}+2^{2 U_2})/4$ stays finite as $r\to 0$. Then, in this limit, the functions describing the static solution behave as 
\bea
&&\lambda_{00}\approx - g(z)\,,\quad \lambda_{11}\approx h(z)\,,\quad \tau\approx g(z) h(z)\equiv f(z)\,,\quad {\tilde U}_+\approx {z\over 2}+c_1\\
&&V_1\approx {z\over 2} + c_2\,,\quad \omega^1_- \approx 1\,,\quad {\tilde V}_1\approx -{z^2\over 4}- z c_2 + c_3\quad \mathrm{if}\quad z\in (\p_{i_1},\p_{i_1+1})\nonumber\\
&& V_1\approx -{z\over 2} + c_2\,,\quad \omega^1_- \approx -1\,,\quad {\tilde V}_1\approx {z^2\over 4}- z c_2 + c_3\quad  \mathrm{if}\quad  z\in (\p_{i_2},\p_{i_2+1})\nonumber\\
\eea
where $f(z)$, $g(z)$ and $h(z)$ are positive functions and $c_1, c_2, c_3$ are constants.\footnote{$f$, $g$, $h$ and $c_i$ depend on the details of the static solution. Their explicit expressions are not relevant for the ensuing argument, and hence will not be given here.} We observe that $\tau$ goes to a non-vanishing function of $z$ as $r\to 0$ and also, from the relations (\ref{kappamess}) and the form of $\omega^1_-$, $V_1$, ${\tilde U}_+$ and ${\tilde V}_1$ given above, that the matrix of 1-forms $\kappa$ is $z$-independent in this limit, and given by
\bea
\kappa=\begin{pmatrix}2 c_1 &0&0\cr 0&-c_1 \pm c_2 +{c^1}_1&\mp 1\cr 0& \pm c_2^2 +c_3& -c_1 \mp c_2 -{c^1}_1 \end{pmatrix}d\phi_-
\label{kappaspacelike}
\eea
where the upper signs apply to the case $z\in (\p_{i_1},\p_{i_1+1})$ and the lower signs to the case $z\in (\p_{i_2},\p_{i_2+1})$ (the constant ${c^1}_1$ has been given in (\ref{c11})).
 It is obvious from the expressions above that all components of the $\chi$ matrix go to finite functions of $z$ in this limit. Thus also the matrix $\chi'$ obtained by applying to $\chi$ the linear transformation (\ref{5Dtrans0}) will
have components that are finite functions of $z$ as $r\to 0$. This proves that $\tau'=-1/{(\chi')^2}_2$ goes to a non-vanishing function of $z$ as $r\to 0$, and thus a region of type 1 is sent into a region of the same type by the transformation (\ref{5Dtrans0}). In this region, the null eigenvector of $G$ for the rotating metric is
\be
v'=\{-(\omega')^a_-,1\}
\ee
where 
\be
(\omega')^a = -{(\kappa')^a}_2\,,\quad \kappa'=(M^T)^{-1}\kappa M^T
\label{kapparotspacelike}
\ee
Since $\kappa$ is a $z$-independent matrix, in this limit, we are guaranteed that the null eigenvector $v'$
is also $z$-independent, within each rod.  Note however that the constants $c_i$ that appear in (\ref{kappaspacelike}) maybe be different for
different rods, so in general one has different null eigenvectors for the different rods. 

\item case 2. Consider taking $z=\p_i$, where $\p_i$ is the separation point between any two rods, and send $r\to 0$. The static solution in this limit behaves as
\be
\lambda_{00}\approx - r g(\p_i)\,,\quad \lambda_{11}\approx h(\p_i)\,,\quad \tau\approx r g(\p_i) h(\p_i)\equiv r f(\p_i)\,,\quad V_1 \approx c
\ee  
and the $\chi$ matrix describing this solution has the form
\be
\chi\approx {1\over f(\p_i)} \begin{pmatrix}
-r f(\p_i) g(\p_i)  &0&0\cr 0&-{c^2\over r}&{c\over r} \cr 0&{c\over r}& -{1\over r}
\end{pmatrix}
\ee
Apply a transformation of the type (\ref{5Dtrans0}), with $M=M_\alpha$, to generate a stationary solution with $\chi$ matrix
\be
\chi'=M_\alpha\chi M^T_\alpha
\ee
One finds that $\tau$ transform as
\be
\tau\to \tau'\approx  {4 f(\p_i)\over (1+c + (1-c) \cosh\alpha)^2} r
\ee
which is of the same form of the original $\tau$. This shows that the points of separation between rods are left unchanged by the transformations (\ref{5Dtrans0}). 
We note that the denominator in $\tau'$ vanishes for some critical value of $\alpha$
\be
\cosh\alpha = {c+1\over c-1}
\ee
At this critical value of $\alpha$ the angular momentum causes the metric to degenerate: This is the analogue of the extremal point of the Myers-Perry geometry.

\item case 3. Consider the limit $r\to 0$ and $z\in (\p_{i_0},\p_{i_0+1})$ for some $i_0$, i.e. $z$ belongs to the interior of a time-like rod. In this case the functions describing the static solution have the form
\be
\lambda_{00}\approx -r^2 g(z)\,,\quad \lambda_{11}\approx h(z)\,,\quad \tau \approx r^2 g(z) h(z) \equiv r^2 f(z)\,,\quad V_1 \approx c
\ee
where $f(z)$, $g(z)$ and $h(z)$ are positive functions and $c$ is a constant. Then the $\chi$ matrix for the static solution has the limit
\be
\chi\approx {1\over f(z)} \begin{pmatrix}
-r^2 f(z) g(z)  &0&0\cr 0&-{c^2\over r^2}&{c\over r^2} \cr 0&{c\over r^2}& -{1\over r^2}
\end{pmatrix}
\ee
Let us act on this matrix with an $\mathrm{SL}(3,\mathbb{R})$ transformation $M$, and let us denote by $\lambda'_{ab}$, $(\omega')^a$ and $\tau'$ the quantities describing the transformed geometry. If we take $M=M_\alpha$, one finds that
\bea
\lambda'&\approx& \lambda(z) \begin{pmatrix} 2 \sinh^2\alpha & \sqrt{2} \sinh 2\alpha\cr  \sqrt{2} \sinh 2\alpha & 4 \cosh^2\alpha\end{pmatrix}\nonumber\\
\tau'&\approx& {4 f(z)\over (1+c + (1-c) \cosh\alpha)^2} r^2
\eea
where $\lambda(z)$ is some function of $z$ that depends on the details of the starting static solution. Thus, for generic values of the ``boost'' parameter
$\alpha$, the stationary solution is still of the form required by case 3. The matrix $\lambda'$ has a null eigenvector given by
\be
v^a_{0,\alpha}=\{1,-{\tanh\alpha\over \sqrt{2}}\}
\ee
We conclude that, excluding degenerate points, the rotating metric has a rod in the region $(\p_{i_0},\p_{i_0+1})$, the same as the original static metric, with a null eigenvector given by
\be
v'_\alpha=\{1,-{\tanh\alpha\over \sqrt{2}},0\}
\ee
The fact that the component of $v'_\alpha$ along $\phi_-$ vanishes means that the horizon angular velocities around the $\psi$ and $\phi$ axes are equal. 
\end{itemize}

To summarize, we have shown that, when acting on static axisymmetric solutions, $\mathrm{SO}(2,1)$ transformations do not change the number or the position of the rods, but
rotate the null eigenvector corresponding to each rod.

\newsection{Myers-Perry from Schwarzschild}
\label{mpsec}
As an example of the formalism developed above, we show that one can derive the doubly spinning Myers-Perry solution~\cite{myersperry} by applying a sequence of  $\mathrm{SO}(2,1)$ transformations on the Schwarzschild metric in five dimensions. With the standard choice of coordinates the Schwarzschild solution is
\be
ds^2 = -Z dt^2 + \frac{dr^2}{Z} + r^2 (d\theta^2 + \sin^2\theta d\phi^2 + \cos^2\theta d\psi^2)
\ee
where
\be
Z= 1-\frac{\mu}{r^2}
\ee
We replace the angular coordinates $\phi$ and $\psi$ by the following
\be
\phi_- = \psi-\phi, \ \xi^1 = \ell (\psi +\phi)
\ee
Here $\ell$ is an arbitrary length scale. Though the following computations are valid for any choice of $\ell$, the intermediate expressions simplify dramatically if we choose
\be
\ell = \frac{\sqrt{\mu}}{2\sqrt{2}}
\ee
We will set $\ell$ to this value for the rest of the computation. The next step in the process is to separate the five dimensional metric into a three dimensional base and a two dimensional fiber. In line with the notation of the previous sections the fiber will be spanned by $t$ and $\xi^1$ while the base metric will be parametrized by $r,\theta$ and $\phi_-$. For the quantities $\lambda_{ab},\tau$ and $ds^2_3$ defined in Eqs. (\ref{dec}), we find
\be
\lambda_{00} =  -Z =, \ \lambda_{11} = \frac{2r^2}{\mu} ,\ \lambda_{01}=0,\ \tau =\frac{2r^2 Z}{\mu}
\ee
\be
ds^2_3 = \frac{2 r^2}{\mu} \left[ dr^2 + (r^2-\mu)\left(d\theta^2 + \frac{1}{4}\sin^2\theta d\phi_-^2\right) \right]
\ee
The gauge potentials $\omega^{0}$ and $\omega^{1}$ are
\be
\omega^{0}=0, \ \omega^{1} = \frac{\sqrt{\mu}}{2\sqrt{2}} \cos 2\theta d\phi_-
\ee
The gauge potential $\omega^{1}$ can be dualized to a scalar by solving the equation below
\be
dV_1 = -\tau \lambda_{11} *_3 d\omega^1
\ee
The solution turns out to be
\be
V_1 = -1 + \frac{2 r^2}{\mu}
\ee
The constant in $V_1$ has been chosen so as to ensure that the $\chi$-matrix to be constructed below has the correct asymptotic form (\ref{b5}). We find that the $\chi$ matrix of the Schwarzschild solution is
\be
\chi_{0} = \left( \begin{array}{ccc} -Z & 0& 0 \\ 0 & -\frac{2 r^2 (Z-1)^2 }{4 \mu Z} & \frac{1}{2} (1+Z^{-1}) \\ 0& \frac{1}{2} (1+Z^{-1}) & -\frac{\mu}{2 r^2 Z}     \end{array} \right)
\ee
As $r \rightarrow \infty$, $\chi_0$ approaches $\eta_5$ defined in Eq. (\ref{b5}). Associated to the $\chi_{0}$ is its dual $\kappa_0$, defined by
\be
d\kappa_0  =  \chi_{0}^{-1}*_{3} d\chi_{0}
\ee
We note that $\chi_0$ is a matrix of scalar fields while $\kappa_{0}$ is a matrix of one forms on the base space. For the Schwarzschild metric, the $\kappa_{0}$ is easily computed to be
\be
\kappa_{0} =-\sqrt{\frac{\mu}{8}}\left( \begin{array}{ccc} 2 & 0& 0 \\ 0 & -1 & 1 \\ 0&  1& -1   \end{array} \right)\cos2\theta d\phi_{-} 
\ee
To apply the first transformation we act with $M_\alpha$ defined in Eqs. (\ref{nabc},\ref{mabc}) on $\chi_{0}$ and $\kappa_0$. The action of $M_\alpha$ on $\chi_0$ and $\kappa_0$ can be found in Eqs. (\ref{5Dtrans0},\ref{krot}). From the transformed $\chi$ one can read off the new $\tau$ and $\lambda_{ab}$ and from the transformed $\kappa_0$ one can read off the new gauge potentials $\omega^0$ and $\omega^1$. On the resulting metric we perform the following set of coordinate transformations
\be
t = \tilde{t} + \frac{1}{4} \sqrt{\mu} \sinh2\alpha (\tilde{\phi} + \tilde{\psi}),\ \xi^1 = \frac{\sqrt{\mu} (1-\sinh^2\alpha)}{2\sqrt{2} } \left(\tilde{\psi } + \tilde{\phi} \right),\ \phi_{-} = \tilde{\psi} -\tilde{\phi}
\ee
\be
r^2 =\frac{1}{1-\sinh^2\alpha} \left( \tilde{r}^2 - \mu \sinh^2\alpha \tanh^2\alpha \right),\ \theta= \ttheta
\ee
In the new coordinates the solution takes the following form
\bea
ds^2 &=& -d\tilde{t}^2 + \frac{M}{\Sigma} (dt - \frac{a}{2} \sin^2\ttheta d\tilde{\phi} - \frac{a}{2} \cos^2\ttheta d\tilde{\psi})^2 + (\tilde{r}^2 + \frac{a^2}{4} ) (\sin^2\ttheta d\tilde{\phi }^2 + \cos^2\ttheta d\tilde{\psi}^2) \nonumber \\
& &+ \frac{\Sigma}{\Delta} d\tilde{r}^2 + \Sigma d\ttheta^2, \label{mpequal}\\
\Sigma &=& \tilde{r}^2 + \frac{a^2}{4},\ \Delta =\tilde{r}^2 \left( 1+ \frac{a^2}{4\tilde{r}^2} \right)^2 -M
\label{mp1}
\eea
where
\be
M = \mu \cosh^2\alpha, \ a= -2\sqrt{\mu} \tanh\alpha
\ee
One can recognize the above solution as the five dimensional Myers-Perry black hole with equal angular momentum parameters $a_1 =a_2=a/2$ and the mass parameter $M$. It is interesting to note that the parametrization of the mass and angular momentum parameters for the final solution is such that $\sinh\alpha =1$ corresponds to the extremal Myers-Perry solution. Furthermore, for all values of $\alpha$ we have $M^2 -  a^2 \geq 0$. This suggests that we cannot ``boost'' the starting Schwarzschild solution to the over rotating regime of the Myers-Perry solution.

\subsection{The Flip}
We have derived the Myers-Perry solution with one independent angular momentum. To derive the general solution, we need to apply the procedure one more time on the second space-like Killing vector. In our notation this corresponds to $\phi_-$ above. This step exemplifies the application of the above formalism to a non-static geometry.

As our starting point we take Eq. (\ref{mpequal}) and perform the following change of coordinates
\be
\tilde{\psi} = \frac{1}{2} \left(\frac{\xi^1}{\ell} + \phi_{+} \right),  \ \tilde{\phi} = -\frac{1}{2} \left(\frac{\xi^1}{\ell} - \phi_{+} \right), \ \tilde{r}^2 = 4 \ell \left( \rho + \frac{M-2a^2}{8 \ell} \right), \tilde{\theta}= 2 \theta
\ee
The $r$ and $\theta$ coordinate transformation is not necessary for the following procedure but it is useful in simplifying the intermediate expressions. On the other hand, the $\tilde{\psi}$ and $\tilde{\phi}$ transformations  are necessary. The procedure is associated to a particular choice of a Killing vector and different choices will lead to physically different solutions.

The Myers-Perry metric with equal angular momentum parameters, given in Eq. (\ref{mp1}),  can be written in the form (\ref{dec}), with
\bea
\lambda_{00}&=&-{8 \ell  \rho-M \over 8\ell \rho+M},\ \lambda_{11} ={(8 \ell  \rho +M)^2 + M a^2 \cos^2\theta \over 8 \ell ^2 (8\ell \rho+M)},\ \lambda_{01}=-{M a \cos\theta \over 2 \ell  (8\ell \rho+M)} \\
\tau &=& {(8 \ell  \rho)^2 + M( a^2 \cos^2\theta - M )\over 8\ell ^2 (8 \ell  \rho +M)},\ \omega^0 = {Ma(8 \ell  \rho + M)  \sin^2\theta \over 2[(8 \ell  \rho)^2 +M(a^2 \cos^2\theta-M)]}, \nonumber \\
 \omega^1 &=& \ell  {[(8 \ell  \rho)^2 -M(M-a^2)]\cos\theta\over (8 \ell  \rho)^2 +M(a^2 \cos^2\theta-M)}\nonumber \\
ds^2_{3}&=&[(8 \ell  \rho)^2 +M (a^2 \cos^2\theta -M)]
\left[ {d\rho^2\over (8 \ell  \rho)^2 -M(M-a^2)}+{d\theta^2\over 64 \ell ^2}\right] \nonumber \\ && +{(8 \ell  \rho)^2 -M(M-a^2)\over 64\ell ^2}\sin^2\theta d\phi^2
\eea
Dualising the $\omega^a$ with respect to the three dimensional metric yields
\be
V_0 = -\frac{M a \cos\theta}{2\ell(8 \ell \rho + M) }, \ V_1 = \frac{8\ell \rho(8 \ell \rho +M) + M a^2 \cos^2\theta }{8 \ell^2 (8 \ell \rho +M)}
\ee
As before, \emph{a priori} the value of $\ell$ is arbitrary, though a judicious choice can simplify the subsequent calculations dramatically. In the current context, it is useful to set
\be
\ell = \frac{\sqrt{M}}{2\sqrt{2} }
\ee
With the above data, the $\chi$ matrix is found to be
\be
\chi = \left( \begin{array}{ccc} -1- \frac{2(\sqrt{8M}\rho +M)}{8\rho^2 -M + a^2 \cos^2\theta} & -\frac{\sqrt{2M}a\cos\theta}{8\rho^2 -M + a^2 \cos^2\theta} &\frac{\sqrt{2M}a\cos\theta}{8\rho^2 -M + a^2 \cos^2\theta} \\ -\frac{\sqrt{2M}a\cos\theta}{8\rho^2 -M + a^2 \cos^2\theta} & - \frac{\sqrt{8M}\rho -M}{8\rho^2 -M+ a^2 \cos^2\theta } & 1- \frac{\sqrt{8M}\rho -M}{8\rho^2-M + a^2\cos^2\theta} \\ \frac{\sqrt{2M}a\cos\theta}{8\rho^2 -M + a^2 \cos^2\theta} &  1- \frac{\sqrt{8M}\rho -M}{8\rho^2-M + a^2\cos^2\theta} & \frac{\sqrt{8M}\rho -M}{8\rho^2-M + a^2 \cos^2\theta} \end{array} \right)
\ee
The corresponding dual matrix of one forms, $\kappa$, is
\be
\kappa = \! \! \left(\!\! \begin{array}{ccc} -\sqrt{\frac{M}{2}} \frac{(8\rho^2 -M + a^2)\cos\theta}{8 \rho^2 -M+ a^2\cos^2\theta} & \frac{a\sqrt{M} (2\sqrt{2} \rho +\sqrt{M} ) \sin^2\theta}{2(8 \rho^2 -M + a^2\cos^2\theta )} & -\frac{a\sqrt{M} (2\sqrt{2} \rho +\sqrt{M} ) \sin^2\theta}{2(8 \rho^2 -M + a^2\cos^2\theta )}  \\
\frac{a\sqrt{M} (2\sqrt{2} \rho -\sqrt{M} ) \sin^2\theta}{2(8 \rho^2 -M + a^2\cos^2\theta )} & \frac{\sqrt{M} \cos\theta}{2\sqrt{2} }\left(1+\frac{a^2\sin^2\theta}{8 \rho^2 -M + a^2 \cos^2\theta} \right) & -\sqrt{\frac{M}{8}} \frac{(8\rho^2 -M + a^2)\cos\theta}{8 \rho^2 -M+ a^2\cos^2\theta}\\
-\frac{a\sqrt{M} (2\sqrt{2} \rho -\sqrt{M} ) \sin^2\theta}{2(8 \rho^2 -M + a^2\cos^2\theta )} & -\sqrt{\frac{M}{8}} \frac{(8\rho^2 -M + a^2)\cos\theta}{8 \rho^2 -M+ a^2\cos^2\theta} &  \frac{\sqrt{M} \cos\theta}{2\sqrt{2} }\left(1+\frac{a^2\sin^2\theta}{8 \rho^2 -M + a^2 \cos^2\theta} \right) \end{array} \!\! \right)
\ee
One can apply a transformation matrix $M_{\tilde\alpha}$ on this data. From the resulting $\chi$ and $\kappa$ we can reconstruct the new metric. Applying the following set of coordinate changes,
\bea
&& t=\tilde{t} + \frac{\sqrt{M}\sinh2\tilde\alpha}{4} (\tilde{\phi} + \tilde{\psi}),\ \xi^1 = \frac{\sqrt{M} (3-\cosh2\tilde{\alpha})}{4\sqrt{2} } (\tilde{\phi} + \tilde{\psi}),\ \phi = (\tilde{\psi} -\tilde{\phi}),\nonumber \\
&& \rho= \frac{\sqrt{2} }{\sqrt{M} (3-\cosh2\tilde\alpha )} \left( \tr^2 + M \tanh^2\tilde\alpha + \frac{a^2}{4} \cosh^2\tilde\alpha -\frac{M}{2} \right), \ \theta = 2\tilde{\theta}   
\eea
transforms the resulting solution to the standard Myers-Perry form with two independent angular momenta.
\bea
ds^2_{MP} &=& -d\tilde{t}^2 + \frac{\tmu}{\Sigma} (dt - a_{1} \sin^2\ttheta d\tilde{\phi} - a_{2} \cos^2\ttheta d\tilde{\psi})^2 + \frac{\Sigma}{\Delta} d\tilde{r}^2 + \Sigma d\ttheta^2 \nonumber \\ &&+(\tilde{r}^2 + a_1^2 ) \sin^2\ttheta d\tilde{\phi }^2+(\tr^2+a_2^2)\cos^2\ttheta d\tilde{\psi}^2 
, \label{mpgeneral}\\
\Sigma &=& \tilde{r}^2 + a_1^2\sin^2\ttheta + a_2^2 \cos^2\ttheta ,\ \Delta =\tilde{r}^2 \left( 1+ \frac{a_1^2}{\tilde{r}^2} \right)\left(1+\frac{a_2^2}{\tilde{r}^2} \right) -\tmu
\eea
where
\be
\tmu = M \cosh^2\tilde\alpha,\ a_1 =   \sqrt{M} \sinh\tilde\alpha +\frac{a}{2}, \ a_2 = \sqrt{M}\sinh\tilde\alpha - \frac{a}{2}
\ee

\newsection{General stationary solutions}
Let us first summarize the results obtained so far. We have started from a five dimensional asymptotically flat static generalized Weyl solution, characterized by some
distribution of rods, two of which are semi-infinite and the remaining $N$ are finite. By applying an $\mathrm{SL}(3,\mathbb{R})$ transformation, we have
generated a family of stationary asymptotically flat solutions. We have shown that, even if the subgroup of $\mathrm{SL}(3,\mathbb{R})$ that preserves asymptotic flatness has 3 parameters, only a 1-parameter family of physically distinct solutions is generated in this way (the one parameter can be chosen to be the ``boost'' $\alpha$ contained in the matrix $M_\alpha$). The solutions so generated have the following properties:
\begin{enumerate}

\item They carry non-vanishing angular momenta, whose value depend on $\alpha$ and on the details of the starting static solution.

\item They have the same number of rods as the static seed solution; also, the position of the rods on the $z$ axis is unchanged by the
transformation.

\item The null eigenvectors corresponding to each rod are rotated with respect to the static solution, by an amount that depends both on $\alpha$ and
on the details of the seed's rod structure. 
\end{enumerate}
This 1-parameter family of solutions certainly does not account for the most general stationary solution with a fixed number of rods, and fixed rod positions.
Let us count the number of independent parameters that such a general solution should have. For a five dimensional asymptotically flat solution, the null eigenvectors corresponding to the two semi-infinite rods can always be chosen to point along the directions $\psi$ and $\phi$. On the other hand, the null eigenvector
corresponding to each finite rod can be parametrized by two independent numbers. Thus for a solution having $N$ finite rods, we expect a total of
$2N$ independent parameters,\footnote{A generic member of this $2N$ dimensional family might represent a geometry with orbifold or Dirac-Misner singularities. These singularities vanish only for some particular orientation of the null eigenvectors corresponding to the rods.} if we keep the position of the rods fixed. The family of solutions we have generated represents a 1-dimensional subspace
of this $2N$-dimensional space of solutions. Can we generate the full $2N$-dimensional space of solutions by repeated applications of  $\mathrm{SO}(2,1)$ transformations?

We have seen in the previous section that for the simplest case of $N=1$ the answer
to this question is positive: if one starts with the static solution with 1 finite rod, i.e. the five dimensional Schwarschild black hole, applies an $\mathrm{SO}(2,1)$
transformation, followed by a ``flip'' (as defined in section 2.4) and then a second $\mathrm{SO}(2,1)$ transformation, one reaches the Myers-Perry
solution with two angular momenta, which is the most general stationary asymptotically flat solution with 1 finite rod. Each action of the $\mathrm{SO}(2,1)$ group only adds one physical degree of freedom, that
one could take to be the ``boost'' parameter $\alpha$.
Further applications of $\mathrm{SO}(2,1)$ transformations do not produce, in this case, any new solutions: this is because there are no more independent parameters to add in the case of 1 finite rod.

We conjecture a generalization of this result to the case of solutions with $N$ finite rods: Start from a static solution with $N$ finite rods. Apply an $\mathrm{SO}(2,1)$ transformation followed by a ``flip'', another $\mathrm{SO}(2,1)$ transformation, and so on. This sequence of transformations is
conjectured not to change the number of rods, nor the position of the rods on the $z$ axis; it  however rotates the null eigenvectors corresponding to each rod. Moreover at each step the $\mathrm{SO}(2,1)$ transformation generates an asymptotically $\mathbb{R}^{4,1}$ solution 
and adds one physical degree of freedom to the family of solutions.\footnote{Because SO(2,1) transformations change the relative orientation between the rods, they might alter the details of the singularity structure of the solution.} After the application of $2 N$ 
$\mathrm{SO}(2,1)$ transformations, one has generated a family of solutions with $2N$ independent parameters, that represents the most general stationary axisymmetric solution, for a fixed number of rods.  Successive iterations of this
procedure do not give rise to any new solutions.

In order to prove the conjecture above, it would be useful to understand the action of  $\mathrm{SO}(2,1)$ transformations on a general stationary solution in more detail. In particular it would be important to know how a sequence of ``flip'' plus $\mathrm{SO}(2,1)$ transformation acts on the rod structure of the solution. This task seems, at first sight, technically challenging because a ``flip'' does not act in any natural way on the matrices
$\chi$ and $\kappa$, on which $\mathrm{SO}(2,1)$ acts linearly. We leave
the analysis of this problem for future work.

Finally, we would like to comment on the matrix $D$, introduced in section 2.3. It can be easily verified that acting with $D$ on a solution does not change
the rod structure, including the (relative) orientation of the null eigenvectors. However the action of $D$ changes the physical properties of the solution: 
it converts a solution that  asymptotically goes to $\mathbb{R}^{4,1}$, to a solution that goes to  $\mathbb{R}^{3,1}\times S^1$, by adding a KK monopole charge to the geometry. In this sense $D$ is the transformation that realizes the 4D-5D connection in the general (non-supersymmetric) case. 
We plan to work out the applications of this observation in future work~\cite{ross,peet}. After this work was completed, we were made aware of~\cite{clementnew} where the relation between the Myers-Perry solution in five dimensions and the Kerr solution in four dimensions has been established.

\section*{Acknowledgments}
We would like to thank Roberto Emparan, Jon Ford, Troels Harmark, Amanda Peet, Andrei Pomeransky, Simon Ross and Yogesh Srivastava for valuable discussions and correspondence. We were supported by NSERC.

\end{document}